\documentclass[twocolumn,amsmath,nofootinbib,%showpacs,%
amssymb]{revtex4}
\usepackage{epsfig}
\usepackage{graphicx}% Include figure files
\usepackage{dcolumn}% Align table columns on decimal point
\usepackage{bm}% bold math
\usepackage{color}

\def\thebiblio#1{
\begin{center}\bf \large References
\end{center}
\list
{[\arabic{enumi}]}{\settowidth\labelwidth{#1.}\leftmargin\labelwidth
 \advance\leftmargin\labelsep
 \usecounter{enumi}}
 \def\newblock{\hskip .11em plus .33em minus -.07em}
 \sloppy
 \sfcode`\.=1000\relax}

\begin{document}
%\begin{titlepage}

\title{Non-minimally coupled vector curvaton}

\author{Konstantinos~Dimopoulos}%
\email{k.dimopoulos1@lancaster.ac.uk}
\affiliation{%
Physics Department, Lancaster University}%
\author{Mindaugas~Kar\v{c}iauskas}%
\email{m.karciauskas@lancaster.ac.uk}
\affiliation{%
Physics Department, Lancaster University}%

\date{\today}% It is always \today, today,
             % but any date may be explicitly specified
%\medskip

\begin{abstract}
It is shown that a massive Abelian vector boson field can generate the 
curvature perturbation in the Universe, when coupled non-minimally to gravity, 
through an $RA^2$ coupling. The vector boson acts as a curvaton field imposing
the curvature perturbation after the end of inflation, without generating a
large-scale anisotropy. The parameter space of the model is fully explored,
obtaining the relevant bounds on the inflation scale and the decay constant
of the vector curvaton.
\end{abstract}

%\end{titlepage}

 % PACS, the Physics and Astronomy
 % Classification Scheme.
%\keywords{Suggested keywords}%Use showkeys class option if keyword
                              %display desired
\maketitle

%\tableofcontents

%\pagebreak

Observations provide strong evidence that the Universe underwent a phase of
inflation in its early history. One of the most important consequences of 
inflation is the generation of the curvature perturbation, which is necessary 
for structure formation and observed through the CMB anisotropy \cite{book}. 
Quantum fluctuations of suitable fields give rise to a flat superhorizon 
spectrum of perturbations through the process of particle production 
\cite{hawking}. Under certain circumstances these perturbations can create the 
curvature perturbation of the Universe. So far only scalar fields have been 
employed for this task. Recently, however, it has been shown that 
Abelian gauge fields can also work \cite{vecurv,sugravec}. Indeed, in 
Ref.~\cite{vecurv} it was shown that, if a vector field obtains a flat 
superhorizon spectrum of perturbations during inflation, it can act as a 
curvaton field \cite{curv} provided, at some point after inflation, its 
mass-square becomes positive and bigger than the Hubble scale. In this case the
vector field condensate oscillates coherently, behaving as pressureless 
{\em isotropic} matter \cite{vecurv}. Thus, it can dominate the radiation 
background without introducing significant anisotropy, imposing thereby its 
own curvature perturbation according to the curvaton mechanism \cite{curv}.
Hence, the mechanism of vector curvaton appears to work using a massive
Abelian gauge field provided an approximately scale-invariant superhorizon
spectrum of its perturbations is created during inflation. In 
Ref.~\cite{vecurv} it was shown that this can be achieved if the effective mass
of the vector field during inflation is \mbox{$m_{\rm eff}^2\approx-2H^2$}, 
where $H$ is the Hubble scale. In this letter we investigate a model with an
Abelian massive gauge field, which is non-minimally coupled to gravity, such 
that the above condition can be satisfied during inflation.

Consider the Lagrangian density:
\begin{eqnarray}
 & {\cal L}=-\frac14 F_{\mu\nu}F^{\mu\nu}+\frac12m^2A_\mu A^\nu+
\frac12\alpha RA_\mu A^\nu, &
\label{L0}
\end{eqnarray}
where \mbox{$F_{\mu\nu}=\partial_\mu A_\nu-\partial_\nu A_\mu$} is the
field strength tensor, $m$ is the bare mass of the gauge field and $R$ is the 
Ricci scalar, with $\alpha$ being a real coupling constant. We assume that
a phase of inflation during the early Universe inflates away its spatial 
curvature. In this case we can employ the spatially flat FRW metric, which 
suggests
\begin{equation}
R=-6\left(\frac{\ddot a}{a}+\frac{\dot a^2}{a^2}\right)=
-6(\dot H+2H^2)\,, 
\label{R}
\end{equation}
where the dot denotes derivative with respect to the cosmic time $t$ and
\mbox{$H\equiv\dot a/a$}, with $a$ being the scale factor of the Universe.
During (quasi)de Sitter inflation \mbox{$H\simeq\,$constant} and
\mbox{$R\simeq -12H^2$}. This means that the effective mass of our vector field
is \mbox{$m_{\rm eff}^2\simeq m^2-12\alpha H^2\simeq\,$constant}.

Now, inflation also homogenises the vector field. Following Ref.~\cite{vecurv},
we can calculate the spectrum of superhorizon perturbations for the vector 
field. We find that the dominant contribution to the power spectrum of the
vector field perturbations is
\begin{equation}
\,\!
{\cal P_A}\simeq\frac{\pi}{1-\cos(2\pi\nu)}\left(\frac{aH}{2\pi}\right)^2
\frac{1}{\Gamma^2(1-\nu)}\left(\frac{k}{2aH}\right)^{3-2\nu}\!\!\!,
\hspace{-1cm}
\label{PA}
\end{equation}
where \mbox{$k\ll aH$} is the comoving momentum scale and 
\begin{equation}
\nu\equiv\sqrt{\frac14-\frac{m_{\rm eff}^2}{H^2}}=
\sqrt{\frac14+12\alpha-\left(\frac{m}{H}\right)^2}.
\label{v}
\end{equation}

The scale dependence of the power spectrum can be parametrised in the usual
way as \mbox{${\cal P_A}\propto k^{n_s-1}$}, so that \mbox{$n_s=1$} corresponds
to a flat spectrum. Comparing this with Eq.~(\ref{PA}) we find that the 
spectral index is
\begin{equation}
n_s-1=3-2\nu\;\Rightarrow\;n_s=4-
\sqrt{1+48\alpha-4\left(\frac{m}{H}\right)^2}.
\label{ns}
\end{equation}

To obtain a scale-invariant spectrum of vector field perturbations
we need
%
%\begin{equation}
%\alpha\approx\frac16\left[1+\frac12\left(\frac{m}{H}\right)^2\right]
%\label{alpha}
%\end{equation}
\begin{eqnarray}
& \alpha\approx\frac16\left[1+\frac12(m/H)^2\right] &
\label{alpha}
\end{eqnarray}
Hence, we see that we need \mbox{$\alpha\gtrsim\frac16$}. If 
\mbox{$m\gtrsim H$} then scale invariance is attained only when $\alpha$
is tuned according to Eq.~(\ref{alpha}). However, if \mbox{$m\ll H$} then
scale-invariance simply requires \mbox{$\alpha\approx\frac16$}. In the latter 
case $m$ and $H$ do not have to balance eachother through the condition in
Eq.~(\ref{alpha}) and can be treated as free parameters. We feel that this is
a more natural setup, so, in the following, we assume 
\mbox{$\alpha\approx\frac16$} unless stated otherwise. Since the
latest observations deviate from exact scale invariance, $\alpha$
should not be exactly equal to 1/6. Indeed, according to the 5-year WMAP 
results \mbox{$n_s=0.960\pm 0.014$} at 1-$\sigma$ \cite{wmap}. This implies
that, when \mbox{$m\ll H$}, we need \mbox{$6\alpha=1.03\pm0.01$}.

To study the evolution of the vector field %during and after inflation 
we consider that, for a homogeneous massive Abelian vector field the temporal 
component $A_t$ is zero \cite{vecurv}, while the spatial components satisfy the
following equation of motion
\begin{eqnarray}
 & \ddot A+H\dot A+\left(m^2+\frac16R\right)A=0\,, & 
\label{EoM}
\end{eqnarray}
where we assume that the homogeneous vector field lies along the $z$-direction
with \mbox{$A_\mu=(0,0,0,A(t)\,)$}. During and after inflation, it is easy to 
show that
\begin{equation}
R=3(3w-1)H^2,
\label{Rw}
\end{equation}
where $w$ is the barotropic parameter of the Universe: \mbox{$w\approx-1$}
[\mbox{$w=\frac13$}] \{\mbox{$w=0$}\} during (quasi)de Sitter inflation
[radiation domination]  \{matter domination\}. Using the above and considering
\mbox{$m\ll H$} we can obtain the following solution for the zero-mode of the
vector field
\begin{equation}
A=W_0a+Ca^{\frac12(3w-1)},
\label{Asolu}
\end{equation}
where $W_0$ and $C$ are constants of integration. Thus, the growing mode for 
the vector field, in all cases, scales as \mbox{$A\propto a$}. 
%The physical interpretation of this result 
This can be understood as follows.

As discussed in Refs.~\cite{vecurv,sugravec} $A_\mu$ is the {\em comoving} 
vector field; with the Universe expansion factored-out. The spatial 
components of the physical vector field, in a FRW geometry are 
\mbox{$W_i\equiv A_i/a$} where $i=1,2,3$. This can be understood just by 
considering the mass term in Eq.~(\ref{L0}), which can be written as
\begin{eqnarray}
 & \frac12m^2A_\mu A^\mu=\frac12m^2(A_t^2-A_iA_i/a^2)\,, & 
\end{eqnarray}
where Einstein summation is assumed. Since the Lagrangian density is a physical
quantity we see that the spatial components of the physical vector field are 
\mbox{$W_i\equiv A_i/a$}. Writing the physical vector field as 
\mbox{$W_\mu=(0,0,0,W(t)\,)$} with \mbox{$W\equiv A/a$}, we can obtain its 
equation of motion from Eq.~(\ref{EoM}) as
\begin{equation}
\ddot W+3H\dot W+m^2W=0\,,
\label{eom}
\end{equation}
which is identical to the one of a massive scalar field and we used 
Eq.~(\ref{Rw}). When \mbox{$m\ll H$} Eq.~(\ref{eom}) has the solution
\begin{equation}
W=W_0+Ca^{\frac32(w-1)},
\label{Wsolu}
\end{equation}
where $W_0$ and $C$ are constants of integration, consistent with
Eq.~(\ref{Asolu}). Thus, %we see that, 
as long as \mbox{$m\ll H$}, the
physical vector field develops a condensate which remains %roughly 
constant
\mbox{$W\simeq W_0$}. This is the physical interpretation of 
\mbox{$A\propto a$}.

We can follow the evolution of the vector field condensate by considering the
energy momentum tensor, which can be written in the form 
\begin{equation}
T^\nu_{\,\mu}={\rm diag}(\rho_A, -p_\perp,-p_\perp,-p_\parallel\,),
\label{T}
\end{equation}
where \cite{mukh}
\begin{eqnarray}
 & \rho_A=\frac12 \dot W^2+\frac12 m^2W^2 &
\label{rhoA}
\end{eqnarray}
and the transverse and longitudinal pressures are \cite{mukh}
\begin{equation}
\,\!
\begin{array}{l}
%p_\perp=\frac56(\dot W^2-m^2W^2)+\frac13(2H\dot W+\dot HW+3H^2W)W\\
p_\perp\!=\!\frac56(\dot W^2\!-\!m^2W^2)\!+\!
\frac13(2H\dot W\!+\!\dot HW\!+\!3H^2W)W\\
\\
%p_\parallel=-\frac16(\dot W^2-m^2W^2)-\frac23(2H\dot W+\dot HW+3H^2W)W
p_\parallel\!=\!-\frac16(\dot W^2\!-\!m^2W^2)\!-\!
\frac23(2H\dot W\!+\!\dot HW\!+\!3H^2W)W.
\end{array}
\label{ps}
\hspace{-1cm}
\end{equation}
Thus, the energy-momentum tensor for the homogeneous 
vector field is, in general, anisotropic because 
\mbox{$p_\parallel\neq p_\perp$}. This is why the vector field cannot be taken 
to drive inflation, for if it did it would generate a substantial large-scale 
anisotropy, which would be in conflict with the isotropy in the CMB. Therefore,
we have to investigate whether, {\em after} inflation, there is a period in 
which the vector field becomes isotropic 
(i.e. \mbox{$p_\perp\approx p_\parallel$}) and can imprint its 
perturbation spectrum onto the Universe.

Considering the growing mode in Eqs.~(\ref{Asolu}) and (\ref{Wsolu}), from
Eqs.~(\ref{rhoA}) and (\ref{ps}) we see that, during and after inflation, 
when \mbox{$m\ll H$}, we have
\begin{eqnarray}
\,\!
\hspace{-.8cm}
& \rho_A\simeq \frac12 m^2W_0^2\;\;{\rm and}\;\;
p_\perp\simeq-\frac12 p_\parallel\simeq\frac12(1-w)H^2W_0^2. &
\label{rA}
\end{eqnarray}
%where \mbox{$W_0=A_1$}. 
Hence, the density of the vector field remains roughly constant,
while %, for \mbox{$w\neq 1$}, 
the vector field condensate remains anisotropic during the hot big bang.

The above are valid under the condition \mbox{$m\ll H$}. However, after the end
of inflation \mbox{$H(t)\propto t^{-1}$}, so there will be a moment when 
\mbox{$m\sim H$}. After this moment, due to Eq.~(\ref{Rw}), the curvature 
coupling becomes negligible and the vector field behaves as a massive 
minimally-coupled Abelian vector boson. As shown in Ref.~\cite{vecurv}, when 
\mbox{$m\gtrsim H$} a massive vector field undergoes (quasi)harmonic 
oscillations of frequency $\sim m$, because the friction term in 
Eqs.~(\ref{EoM}) and (\ref{eom}) becomes negligible. In this case, 
on average over many oscillations, it has been shown that
\mbox{$\overline{\dot W^2}\approx m^2\overline{W^2}$} \cite{vecurv}. 
%Consequently
Hence, Eqs.~(\ref{rhoA}) and (\ref{ps}) become
\begin{equation}
%\,\!
%\hspace{-.5cm}
\begin{array}{l}
\rho_A\simeq m^2\overline{W^2} \quad{\rm and}\\
\\
p_\perp\simeq-\frac12 p_\parallel\simeq
\frac23mH\left[1+\frac34(1-w)(H/m)\right]\overline{W^2}. 
\end{array}
\label{rposc}
\end{equation}
The effective barotropic parameters of the vector field are
\begin{eqnarray}
\hspace{-.5cm} & 0<w_\perp\simeq-\frac12 w_\parallel=\frac23
\left[1+\frac34(1-w)\left(\frac{H}{m}\right)\right]\left(\frac{H}{m}\right)
\ll 1, &
\label{ws}
\end{eqnarray}
where \mbox{$w_\perp=p_\perp/\rho_A$} and 
\mbox{$w_\parallel=p_\parallel/\rho_A$}.
By virtue of the condition \mbox{$m\gg H$}, we see that, after the onset 
of the oscillations, \mbox{$w_\perp,w_\parallel\rightarrow 0$}. 
This means that the oscillating massive vector field
behaves as pressureless {\em isotropic} matter, which can dominate the Universe
without generating a large-scale anisotropy. Moreover, its density can be shown
to decrease as \mbox{$\rho_A\propto a^{-3}$} (like dust) as expected
\cite{vecurv}. 
Thus, if the Universe is radiation dominated, \mbox{$\rho_A/\rho\propto a$} 
while oscillations occur, so the field has a chance to
dominate the Universe and imprint its curvature perturbation according to the
curvaton scenario~\cite{curv}.

At the onset of the oscillations we have
\begin{equation}
\Omega\equiv\frac{\rho_A}{\rho}\sim\left(\frac{W_0}{m_P}\right)^2,
\label{W}
\end{equation}
where we used the flat Friedman equation \mbox{$\rho=3m_P^2H^2$} with
\mbox{$m_P=2.4\times 10^{18}\,$GeV} being the reduced Planck mass.
To avoid excessive anisotropy the density of the vector field must be 
subdominant before the onset of oscillations, which means that 
\mbox{$W_0<m_P$}.

Let us assume that inflation is driven by some inflaton field, which after 
inflation ends, oscillates around its VEV until its decay into a thermal bath 
of relativistic particles at reheating. In this scenario the Universe is matter
dominated (by %massive 
inflaton particles) until reheating. Using the above 
findings we can estimate the Hubble scale when the vector
field dominates the Universe as 
\begin{equation}
H_{\rm dom}\sim\min\{m,\Gamma\}\left(\frac{W_0}{m_P}\right)^4,
\label{Hdom}
\end{equation}
where $\Gamma$ is the decay rate of the inflaton field. If inflation gives away
directly to a thermal bath of particles then we have prompt reheating and 
\mbox{$\Gamma\rightarrow H_*$}, where $H_*$ is the Hubble scale of inflation.
There is a chance, however, that the vector field itself decays before it
dominates the Universe while still being able to act as curvaton. In this case,
the density ratio of the vector field at decay is
\begin{equation}
\Omega_{\rm dec}\sim
\left(\frac{\min\{m,\Gamma\}}{\Gamma_A}\right)^{1/2}
\left(\frac{W_0}{m_P}\right)^2,
\label{Wdec}
\end{equation}
where $\Gamma_A$ is the vector field decay rate. 

According to the curvaton scenario the  gauge invariant comoving
curvature perturbation is \cite{curv}
\begin{equation}
\zeta\sim\Omega_{\rm dec}\zeta_A\;,
\label{z}
\end{equation}
where $\zeta_A$ is the curvature perturbation attributed to the curvaton field.
In a foliage of spacetime of spatially flat hypersurfaces \cite{curv}
\begin{equation}
\zeta_A=-H\,\frac{\delta\rho_A}{\dot\rho_A}=
\frac13\left.\frac{\delta\rho_A}{\rho_A}\right|_{\rm dec}\,,
\label{zA0}
\end{equation}
where we used that the vector field decays after the onset of the oscillations
%(i.e. \mbox{$\Gamma_A<m$}) 
in which case \mbox{$\rho_A\propto a^{-3}$}. Note 
that, since $\zeta_A$ is determined by the fractional perturbation of the 
field's density, which is a scalar quantity, the perturbation $\zeta_A$ is 
scalar and not vector in nature. %, despite being due to a vector field. 

Now, since 
Eq.~(\ref{eom}) is a linear differential equation, $W$ and its perturbation 
$\delta W$ satisfy the same equation of motion. Therefore, they evolve in the 
same way, which means that %the fractional perturbation 
$\delta W/W$ remains
constant, before and after the onset of oscillations. 
%The same is true for
%$W$ and its perturbation $\delta W$ which satisfy Eq.~(\ref{eom}). 
%Hence, we
%have
%%
%\begin{equation}
%\frac{\delta A}{A}=\frac{\delta W}{W}\simeq{\rm constant}.
%\end{equation}
As shown in Ref.~\cite{vecurv}, during the (quasi)harmonic oscillations of the 
massive vector field, \mbox{$\rho_A=m^2\hat W^2$}, where 
$\hat W$ is the amplitude of the oscillating physical vector field. From 
the above we obtain
\begin{equation}
\zeta_A=\frac23\left.\frac{\delta \hat W}{\hat W}\right|_{\rm dec}
=\frac23\left.\frac{\delta W}{W}\right|_{\rm osc}
=\frac23\left.\frac{\delta W}{W}\right|_*,
\label{zA1}
\end{equation}
where `osc' denotes the onset of oscillations and
the star denotes the time when cosmological scales exit the horizon 
during inflation. 

If \mbox{$m\ll H$} during inflation the physical vector field
(not being conformally invariant) undergoes particle production and obtains 
an approximately flat superhorizon spectrum of perturbations, as shown.
Indeed, under the condition in Eq.~(\ref{alpha}), \mbox{$\nu\approx\frac32$} 
and Eq.~(\ref{PA}) gives \cite{vecurv,sugravec}
\begin{equation}
\sqrt{\cal P_A}\approx
%\frac{aH}{2\pi}
aH/2\pi
\;\Rightarrow\;
\sqrt{\cal P_W}\equiv\sqrt{\cal P_A}/a=
%\frac{H}{2\pi}
H/2\pi
\,,
\label{dW}
\end{equation}
i.e. given by the Hawking temperature for de Sitter space, exactly as is the 
case of light scalar fields \cite{hawking}. Hence, from Eqs.~(\ref{zA1}) and
(\ref{dW}) we can write
\begin{equation}
\zeta_A=\frac{H_*}{3\pi W_0}\,.
\label{zA}
\end{equation}
Thus, from the above and Eq.~(\ref{z}) we obtain
\begin{equation}
\zeta\sim\Omega_{\rm dec}
%\frac{H_*}{W_0}.
\,H_*/W_0\;.
\label{zeta}
\end{equation}
Using this, Eqs.~(\ref{Hdom}) and (\ref{Wdec}), after some algebra, we get
\begin{equation}
\frac{H_*}{m_P}\sim\frac{\zeta}{\sqrt{\Omega_{\rm dec}}}
\left(\frac{\max\{H_{\rm dom},\Gamma_A\}}{\min\{m,\Gamma\}}\right)^{1/4}.
\label{H*}
\end{equation}
The Hot Big Bang has to begin before nucleosynthesis (which occurs at 
temperature \mbox{$T_{\rm BBN}\sim 1\,$MeV}). Hence, 
\mbox{$\max\{H_{\rm dom},\Gamma_A\}\gtrsim T_{\rm BBN}^2/m_P$}.
Using this and also \mbox{$\min\{m,\Gamma\}\lesssim H_*$},
we obtain the bound
\begin{equation}
\,\!
H_*\gtrsim\zeta^{4/5}
\,\Omega_{\rm dec}^{-2/5}
(T_{\rm BBN}^2m_P^3)^{1/5}
\Rightarrow
V_*^{1/4}\gtrsim 10^{12}\,{\rm GeV},
\hspace{-1cm}
\label{Hbound}
\end{equation}
where we used that \mbox{$\Omega_{\rm dec}\lesssim 1$} and
\mbox{$\zeta=4.8\times 10^{-5}$} from COBE observations.
This is similar to the case of a scalar field curvaton \cite{lyth}. 
%However, there are important differences, the most obvious one being that,
%for a scalar curvaton, the field remains frozen before the onset of the 
%oscillations with constant density so that its density parameter $\Omega_\phi$
%grows as \mbox{$\Omega_\phi\propto a^{3(1+w)}$}. This difference does not 
%manifest itself in the lower bound on the inflation scale because the bound
%corresponds to prompt reheating with immediate onset of oscillations 
%\mbox{$m\sim H_*$}. 

Another bound on the inflation scale is obtained by considering that
\mbox{$\Gamma_A\sim g^2m$}, where $g$ is the vector field coupling to its 
decay products, for which \mbox{$g\gtrsim m/m_P$} due to gravitational decay. 
Thus, \mbox{$\max\{H_{\rm dom},\Gamma_A\}\gtrsim g^2m$}. 
Combining with Eq.~(\ref{H*}) we obtain the bound
\begin{equation}
H_*\gtrsim
%\frac{\zeta}{\sqrt{\Omega_{\rm dec}}}
\zeta\,\Omega_{\rm dec}^{-1/2}
(m_Pm)^{1/2}
\;\Rightarrow\;V_*^{1/4}\gtrsim 10^{11}\,{\rm GeV}\,,
\label{Vbound}
\end{equation}
where we took \mbox{$m\gtrsim 1\,$TeV}. 
%and also 
%\mbox{$g\gtrsim m/m_P$}, due to gravitational decay. 

Finally, an upper bound on inflation scale can be obtained by combining 
Eq.~(\ref{zeta}) with the requirement \mbox{$W_0<m_P$}, thereby finding
\begin{equation}
H_*<
%\frac{\zeta m_P}{\Omega_{\rm dec}}
\zeta m_P/\Omega_{\rm dec}
\;\Rightarrow\;
V_*^{1/4}<
%10^{16}\Omega_{\rm dec}^{-1/2}\,{\rm GeV}\,.
10^{17}\,{\rm GeV}\,,
\label{upbound}
\end{equation}
where we considered that \mbox{$\Omega_{\rm dec}\gtrsim 10^{-2}$}, 
in order to avoid excessive non-Gaussianity in the CMB \cite{curv}. 

%The fate of the vector field condensate needs to be carefully considerred
%throughout its evolution, if the above mechanism is to be realised. 
We also need to consider the hazardous possibility of the
thermal evaporation of the vector field condensate. Were this 
to occur, all memory of the superhorizon spectrum of perturbations would 
be erased. Considering that the scattering rate of the massive vector bosons 
with the thermal bath is \mbox{$\Gamma_{\rm sc}\sim g^4T$} we can
obtain a bound such that the condensate does not evaporate before the vector 
field decays. Since \mbox{$\Gamma_{\rm sc}/\Gamma_A\propto a^{-1}$}, 
we need to enforce this bound at the onset of the oscillations, when
\mbox{$\Gamma_{\rm sc}\sim g^4\sqrt{m_Pm}$}. Hence, the range for $g$ is 
\begin{equation}
\frac{m}{m_P}\lesssim g\lesssim \left(\frac{m}{m_P}\right)^{1/4}\,,
\label{grange}
\end{equation}
where the lower bound is due to gravitational decay. Note that, in the case
when the vector curvaton dominates the Universe before its decay the condensate
may not evaporate even if the above upper bound is violated. This is because, 
after domination, the density of the thermal bath is exponentially smaller 
than $\rho_A$ by a factor of $(H_{\rm dom}/H)^{2/3}$. Moreover, even if it 
does evaporate the condensate has already imprinted $\zeta_A$ onto the 
Universe at domination rendering the evaporation bound irrelevant.

The above lower bounds on $H_*$ can be substantially relaxed
by employing the so-called mass increment mechanism
according to which, the vector
field obtains its bare mass at a phase transition (denoted by `pt') with 
\mbox{$m/H_{\rm pt}\gg 1$}. The mechanism was firstly introduced for the scalar
curvaton in Ref.~\cite{low} and has been already implemented in the vector 
curvaton case in Ref.~\cite{sugravec}. 

\begin{figure}[htbp]
\hspace{5cm}\epsfig{file=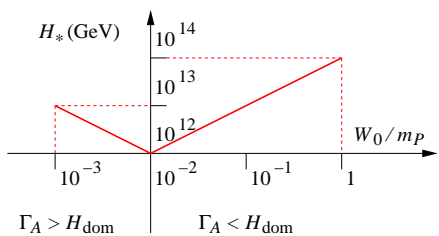, height=3cm}\caption{%
Parameter space for $H_*$ and $W_0$ in our example.
}\label{fig}
\end{figure}

To illustrate our findings let us consider a specific example. Let us choose
\mbox{$m\sim 10\,$TeV} and also \mbox{$\Gamma_A\sim 10^{-10}\,$GeV} such that
the temperature at the vector field decay is \mbox{$T_{\rm dec}\sim 10\,$TeV}.
Such a particle may be potentially observable in the LHC. These values suggest
\mbox{$g\sim 10^{-7}$}, which lies comfortably within the range in 
Eq.~(\ref{grange}). For the decay rate of the inflaton let us chose
\mbox{$\Gamma\sim 10^{-2}\,$GeV} so that the reheating temperature satisfies 
the gravitino overproduction constraint 
\mbox{$T_{\rm reh}\sim\sqrt{m_P\Gamma}\sim 10^8\,$GeV}. 
Assume at first that 
the vector curvaton decays before domination \mbox{$\Gamma_A\geq H_{\rm dom}$}.
Then Eq.~(\ref{H*}) reduces to 
\mbox{$H_*/m_P\sim 10^{-2}\zeta/\sqrt{\Omega_{\rm dec}}$}. Using this and
Eq.~(\ref{zeta}) we get \mbox{$W_0/m_P\sim 10^{-2}\sqrt{\Omega_{\rm dec}}$}.
Hence, the lowest value for the inflation Hubble scale is 
\mbox{$H_*\gtrsim 10^{12}\,$GeV}. It can be readily 
checked that the bound in Eq.~(\ref{Vbound}) is weaker by a factor $10^{-5}$.
Suppose now that the vector curvaton dominates before its decay
\mbox{$\Gamma_A<H_{\rm dom}$}. Using Eq.~(\ref{Hdom}) we get 
\mbox{$W_0>10^{-2}m_P$}, while Eq.~(\ref{H*}) suggests 
\mbox{$H_*\sim\zeta W_0$}. Taking into account the bound \mbox{$W_0<m_P$}, we
find that the maximum value for the Hubble scale is \mbox{$H_*<10^{14}\,$GeV}. 
The relation between $H_*$ and $W_0$ in both cases is depicted in 
Fig.~\ref{fig}.

Let us consider now the case when \mbox{$\alpha\not\approx\frac16$}.
If \mbox{$\alpha={\cal O}(1)$} then, according to Eq.~(\ref{alpha}), a scale
invariant spectrum is possible only if \mbox{$m\sim H_*$}. Hence, the 
oscillations begin immediately after the end of inflation. With this in mind
the previous analysis remains valid. In particular, the bound in 
Eq.~(\ref{Hbound}) remains the same. However, the bound in Eq.~(\ref{Vbound})
becomes much more stringent:
\begin{equation}
H_*\gtrsim\zeta^2m_P\;\Rightarrow\;V_*^{1/4}\gtrsim 10^{14}\,{\rm GeV}\,.
\end{equation}
Hence, in view of
%from the above and 
Eq.~(\ref{upbound}), we see that the inflation energy 
scale is constrained near that of grand unification.

In summary, we have discussed a concrete model which generates the curvature 
perturbation in the Universe with a single massive Abelian vector boson field 
non-minimally coupled to gravity through an $RA^2$ coupling. The vector field
can act as a curvaton, imposing its scalar perturbation spectrum well after the
end of inflation without introducing a large-scale anisotropy. We have shown 
that there is ample parameter space for the model to work by considering 
all relevant constraints in the cosmology. The VEV of the vector curvaton is
zero, which means that it does not violate Lorentz invariance in the vacuum.
Our model does not need to rely on scalar fields at all since inflation might
take place due to purely geometrical effects, such as in $f(R)$-gravity models
\cite{fR} (e.g. $R^2$-inflation \cite{staro}).
The remaining challenge is to realise our mechanism in the context of a 
realistic setup beyond the standard model \cite{sahu}.

Recently, vector fields have been employed to drive inflation \cite{mukh}
(see also Ref.~\cite{ford}). To avoid a large-scale 
anisotropy the authors of Ref.~\cite{mukh} introduce a large number of vector 
fields randomly orientated in space. However, they do not consider the 
generation of curvature perturbations, which could proceed along
the lines of this work, albeit introduced during and not after inflation.
%\cite{future}. 

%Finally, vector fields have also been considered to account for dark energy
%\cite{DE}.

\begin{thebiblio%graphy
}{99}

\bibitem{book}
A.~R.~Liddle and D.~H.~Lyth,
{\em Cosmological Inflation and Large Scale Structure},
(Cambridge University Press, Cambridge U.K., 2000).

\bibitem{hawking}
  G.~W.~Gibbons and S.~W.~Hawking,
  %``Cosmological Event Horizons, Thermodynamics, And Particle Creation,''
  Phys.\ Rev.\  D {\bf 15} (1977) 2738.
  %%CITATION = PHRVA,D15,2738;%%

\bibitem{vecurv}
K.~Dimopoulos,
  %``Can a vector field be responsible for the curvature perturbation in the
  %universe?,''
  Phys.\ Rev.\  D {\bf 74} (2006) 083502.
%  [arXiv:hep-ph/0607229].
  %%CITATION = PHRVA,D74,083502;%%

\bibitem{sugravec}
K.~Dimopoulos,
  %``Supergravity inspired Vector Curvaton,''
  Phys.\ Rev.\  D {\bf 76} (2007) 063506.
%  [arXiv:0705.3334 [hep-ph]].
  %%CITATION = PHRVA,D76,063506;%%

\bibitem{curv}
D.~H.~Lyth and D.~Wands,
%``Generating the curvature perturbation without an inflaton,''
Phys.\ Lett.\ B {\bf 524} (2002) 5;
%[arXiv:hep-ph/0110002].
%%CITATION = HEP-PH 0110002;%%
T.~Moroi and T.~Takahashi,
%``Effects of cosmological moduli fields on 
%cosmic microwave background,''
Phys.\ Lett.\ B {\bf 522}, 215 (2001)
[Erratum-ibid.\ B {\bf 539}, 303 (2002)];
%[arXiv:hep-ph/0110096].
%%CITATION = HEP-PH 0110096;%%
K.~Enqvist and M.~S.~Sloth,
%``Adiabatic CMB perturbations in 
%pre big bang string cosmology,''
Nucl.\ Phys.\ B {\bf 626} (2002) 395;
%[arXiv:hep-ph/0109214].
%%CITATION = HEP-PH 0109214;%%
%  [arXiv:hep-ph/0110002];
  %%CITATION = PHLTA,B524,5;%%
K.~Dimopoulos and D.~H.~Lyth,
  %``Models of inflation liberated by the curvaton hypothesis,''
  Phys.\ Rev.\  D {\bf 69} (2004) 123509.
%  [arXiv:hep-ph/0209180].
  %%CITATION = PHRVA,D69,123509;%%

\bibitem{wmap}
E.~Komatsu {\it et al.}  [WMAP Collaboration],
  %``Five-Year Wilkinson Microwave Anisotropy Probe (WMAP\altaffilmark 1 )
  %Observations:Cosmological Interpretation,''
  arXiv:0803.0547 [astro-ph].
  %%CITATION = ARXIV:0803.0547;%%

\bibitem{mukh}
A.~Golovnev, V.~Mukhanov and V.~Vanchurin,
  %``Vector Inflation,''
  arXiv:0802.2068 [astro-ph].
  %%CITATION = ARXIV:0802.2068;%%

\bibitem{lyth}
D.~H.~Lyth,
  %``Can the curvaton paradigm accommodate a low inflation scale,''
  Phys.\ Lett.\  B {\bf 579} (2004) 239.
%  [arXiv:hep-th/0308110].
  %%CITATION = PHLTA,B579,239;%%

\bibitem{low}
K.~Dimopoulos, D.~H.~Lyth and Y.~Rodriguez,
  %``Low scale inflation and the curvaton mechanism,''
  JHEP {\bf 0502} (2005) 055.
%  [arXiv:hep-ph/0411119].
  %%CITATION = JHEPA,0502,055;%%

\bibitem{fR}
S.~Nojiri and S.~D.~Odintsov,
  %``Modified gravity with negative and positive powers of the curvature:
  %Unification of the inflation and of the cosmic acceleration,''
  Phys.\ Rev.\  D {\bf 68} (2003) 123512.
%  [arXiv:hep-th/0307288].
  %%CITATION = PHRVA,D68,123512;%%

\bibitem{staro}
M.~B.~Mijic, M.~S.~Morris and W.~M.~Suen,
  %``The R**2 Cosmology: Inflation Without A Phase Transition,''
  Phys.\ Rev.\  D {\bf 34} (1986) 2934;
  %%CITATION = PHRVA,D34,2934;%%
B.~Whitt,
  %``Fourth Order Gravity As General Relativity Plus Matter,''
  Phys.\ Lett.\  B {\bf 145} (1984) 176;
  %%CITATION = PHLTA,B145,176;%%
A.~A.~Starobinsky,
  %``A new type of isotropic cosmological models without singularity,''
  Phys.\ Lett.\  B {\bf 91} (1980) 99.
  %%CITATION = PHLTA,B91,99;%%

\bibitem{sahu}
K.~Dimopoulos, M.~Kar\v{c}iauskas and N.~Sahu, in preparation.

\bibitem{ford}
L.~H.~Ford,
%``Inflation Driven By A Vector Field,''
Phys.\ Rev.\ D {\bf 40} (1989) 967;
%%CITATION = PHRVA,D40,967;%%
C.~M.~Lewis,
%``Vector inflation and vortices,''
Phys.\ Rev.\ D {\bf 44} (1991) 1661;
%%CITATION = PHRVA,D44,1661;%%
%
T.~Koivisto and D.~F.~Mota,
  %``Accelerating Cosmologies with an Anisotropic Equation of State: Vector
  %Fields, Modified Gravity and Astrophysical Constraints,''
  arXiv:0801.3676 [astro-ph].
  %%CITATION = ARXIV:0801.3676;%%

%\bibitem{future}
%K.~Dimopoulos, in preparation.

\end{thebiblio}%graphy}

%\pagebreak

\end{document}